\documentclass[useAMS,usenatbib]{mn2e}
\usepackage{epsfig}
\usepackage{amsmath, amssymb,bm}

\title[Dead Zones around Young Stellar Objects]{
 Dead Zones around Young Stellar Objects: FU Orionis Outbursts and Transition Discs }

\author[R. G. Martin et al.]{Rebecca G. Martin$^1$, Stephen
  H. Lubow$^1$, Mario Livio$^1$, and J. E. Pringle$^{1,2}$\\ $^1$Space
 Telescope Science Institute, 3700 San Martin Drive, Baltimore, MD
  21218, USA \\ $^2$Institute of Astronomy, Madingley Road,
  Cambridge, CB3 0HA, UK \\}

\begin{document}

\date{}

\pagerange{\pageref{firstpage}--\pageref{lastpage}} 
\pubyear{2010}
\maketitle

\label{firstpage}

\begin{abstract}
We perform global time-dependent simulations of an accretion disc
around a young stellar object with a dead zone (a region where the
magneto-rotational instability cannot drive turbulence because the
material is not sufficiently ionised). For infall accretion rates on
to the disc of around $10^{-7}\,\rm M_\odot\,yr^{-1}$, dead zones
occur if the critical magnetic Reynolds number is larger than about
$10^4$. We model the collapse of a molecular gas cloud. At early times
when the infall accretion rate is high, the disc is thermally ionised
and fully turbulent. However, as the infall accretion rate drops, a
dead zone may form if the critical magnetic Reynolds number is
sufficiently large, otherwise the disc remains fully turbulent. With a
dead zone the disc can become unstable to the gravo-magneto
instability. The mass of the star grows in large accretion outbursts
that may explain FU Orionis events. At late times there is not
sufficient mass in the disc for outbursts to occur but the dead zone
becomes even more prominent as the disc cools. Large inner dead zones
in the later stages of disc evolution may help to explain observations
of transition discs with an inner hole.
\end{abstract}


\section{Introduction}

The core of a protostellar molecular cloud collapses to form a
low-mass protostar with a disc on a timescale of a few $10^5\,\rm yr$
\citep{shu87}.  Angular momentum is transported outwards through the
disc by turbulence, thus allowing material to accrete on to the
central star \citep{pringle81}. Turbulence in accretion discs may be
driven by the magneto-rotational instability \citep[MRI,][]{balbus91}
but can be suppressed by a low ionisation fraction
\citep{gammie96,gammie98}. The inner parts of the disc may be hot
enough to be thermally ionised and fully MRI active
(turbulent). Further out, where the disc is cooler, a dead zone forms
at the midplane. In this region the MRI cannot drive turbulence. The
surface layers of the disc are ionised by cosmic rays or X-rays from
the star and may remain turbulent
\citep[e.g.][]{sano00,matsumura03}. A second form of turbulence may
also be driven by gravitational instability if the disc becomes
massive enough \citep{paczynski78,lodato04}.

Previously, time-dependent numerical disc simulations have used a
simple description for the extent of the dead zone
\citep{armitage01,zhu10a,zhu10b}. They assume that cosmic rays
penetrate a constant surface density in the disc surfaces and are
attenuated exponentially with a stopping depth of less than $100\,\rm
g\,cm^{-2}$ \citep[e.g.][]{umebayashi81,zhu09, terquem08,matsumura09}.
A more realistic way to determine the dead zone extent is with a
magnetic Reynolds number.  Turbulence is suppressed if the magnetic
Reynolds number is less than some critical value $Re_{\rm M}<Re_{\rm
  M,crit}$ \citep{fleming00}. However, as we discussed in a preceding
paper \citep[][MLLP1]{martin12}, the critical value is uncertain. The
magnetohydrodynamical (MHD) simulations of \cite{fleming00} suggest it
is of the order of $10^4$ if there is no magnetic flux through the
disc, but it may be close to 100 if there is a magnetic flux. The
linear stability analysis of \cite{wardle99} and \cite{balbus01}
suggest it may be as small as $1$, but conditions for linear stability
are not necessarily the same as those for maintaining turbulence
\citep{balbus00}. In MLLP1 we showed that a constant surface density
in the active layer may only be a good approximation for a disc with a
low critical magnetic Reynolds number of around 1. Even then, the
surface density of the layer may be much larger than the constant
values that have been assumed, that were less than about $200\,\rm
g\,cm^{-2}$.

The luminosity of the protostar depends upon the accretion rate
through the disc. The typical bolometric luminosity of protostars is
much smaller than would be expected from the observed infall rate and
duration of the protostellar phase \citep{kenyon90}.  Time-dependent
accretion is thought to be a solution to this luminosity problem.
Observational evidence for this scenario has been seen in FU Orionis
systems that show outbursts with a peak accretion rate of around
$10^{-4}\,\rm M_\odot\,yr^{-1}$ lasting a timescale of years to
decades \citep{herbig77,hartmann96}. A plausible explanation for FU
Orionis outbursts is a disc instability resulting from the formation
of a dead zone, known as the gravo-magneto instability.  The MRI
turbulent active layers of the disc supply material to the dead zone
that builds up until it becomes self-gravitating. Self-gravity drives
a second type of turbulence that can dissipate sufficient heat for the
MRI to be triggered. The now very high surface density disc flows on
to the central star in an accretion outburst. The disc then cools, the
dead zone re-forms and the cycle repeats \citep{armitage01,
  zhu09}. This process can be understood as a limit cycle
\citep{martin11a}. There are two steady state solutions, one with a
fully turbulent disc and a second that is self-gravitating. There is a
range of accretion rates for which no steady state disc solution
exists. The outbursts can be explained as transitions between the two
steady solutions in a state diagram that plots the accretion rate
through the disc against the surface density at a fixed radius.

In the later stages of disc evolution, the infall accretion
slows. There is not enough mass flowing through the disc for the
gravo-magneto instability to operate on a reasonable timescale yet
there may still be a dead zone present. Observations of protostellar
discs show that about five to ten percent are transition discs
\citep{skrutski90}.  These have no significant near-IR excess compared
to younger discs and are thought to have a hole in the inner regions
up to sizes of several tens of AU
\citep[e.g.][]{calvert02,forrest04}. There is still some gas accretion
on to the star but no dust is observed there. There have been several
models suggested to explain these discs such as planets
\citep{rice03}, disc evolution \citep{dullemond05}, disc clearing
\citep{clarke01,alexander06} and the presence of companions
\citep{jensen97}.  \cite{alexander09} suggest that young discs contain
a planet whereas the older ones may be undergoing disc clearing. We
consider an additional possibility that transition discs may contain a
large dead zone that causes a sharp transition in the surface density
of the disc and allows only a small amount of accretion on to the
central star.

In Section~\ref{model} we describe the layered disc model. In
Section~\ref{steady} we find fully turbulent steady state disc
solutions with a constant infall accretion rate and consider
parameters for which a dead zone can form. In Section~\ref{time} we
solve the time-dependent layered accretion disc equations and find the
evolution of a system with a dead zone and a constant infall accretion
rate. In Section~\ref{dec} we model the infall accretion rate from a
collapsing gas cloud and consider disc parameters necessary for the
gravo-magneto instability to explain observed FU Orionis outbursts. In
Section~\ref{trans} we discuss implications of a dead zone model for
transition disc observations.

\section{Layered Disc Model}
\label{model}

\begin{figure*}
\includegraphics[width=8.4cm]{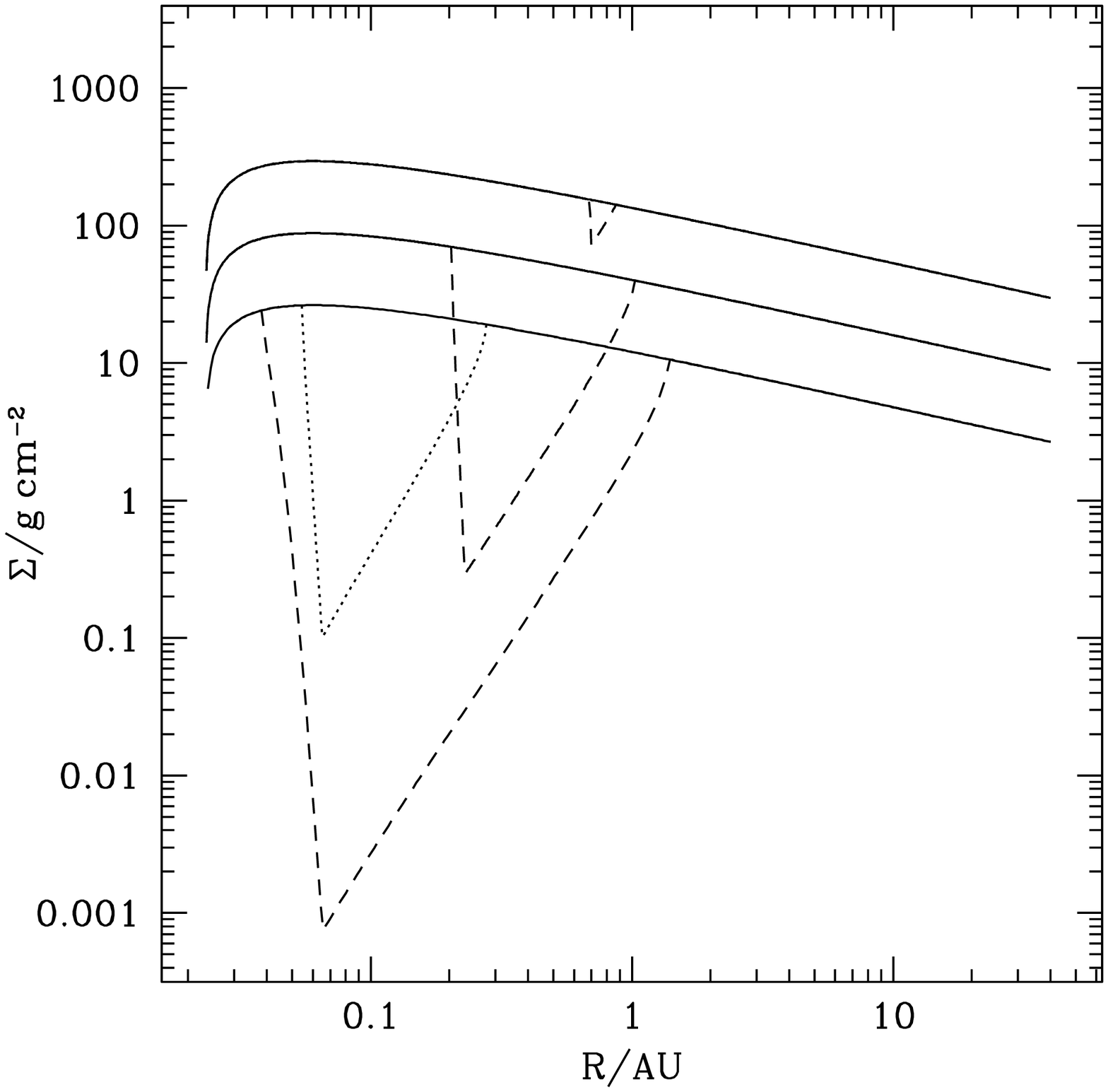}
\includegraphics[width=8.4cm]{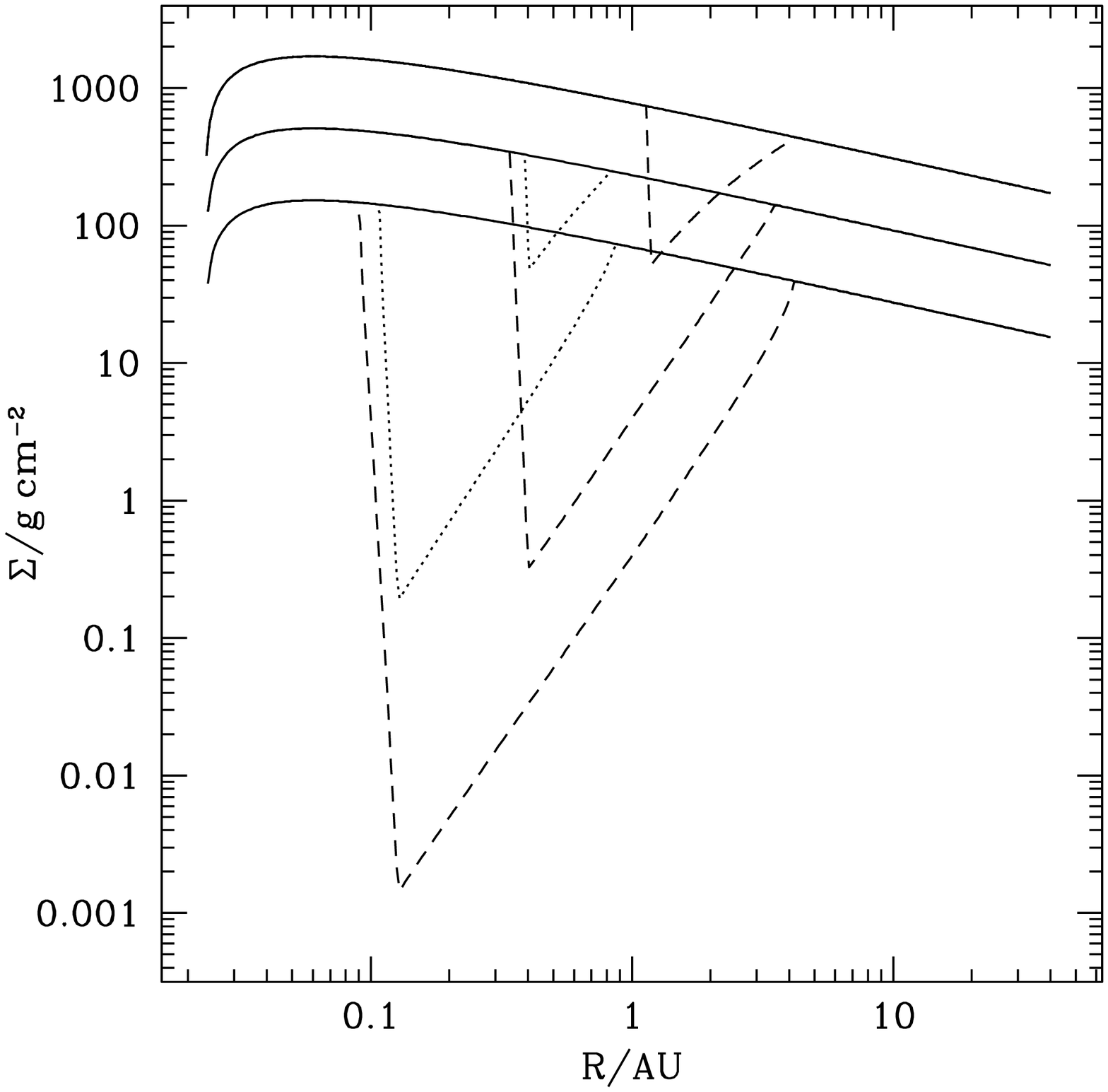}
\caption{Steady state disc solutions that are fully turbulent with
  $\alpha=0.1$ (left) and $\alpha=0.01$ (right). The solid lines show
  the total surface density.  The upper lines in each plot show a disc
  with an infall accretion rate of $\dot M_{\rm infall}=10^{-6}\,\rm
  M_\odot\,yr^{-1}$ the middle lines $\dot M_{\rm infall}=10^{-7}\,\rm
  M_\odot\,yr^{-1}$ and the lower lines $\dot M_{\rm
    infall}=10^{-8}\,\rm M_\odot\,yr^{-1}$. The dashed lines show the
  surface density that is turbulent for $Re_{\rm M,crit}=10^4$ and the
  dotted lines for $Re_{\rm M,crit}=10^3$. The dotted lines are not
  visible in all cases because the disc is fully turbulent and the
  active surface density is equal to the total surface
  density. However, if a dead zone exists in the disc, then the disc
  can not be in steady state.  }
\label{layer}
\end{figure*}

In an accretion disc, material orbits the central mass, $M$, at
Keplerian velocity at radius $R$ with angular velocity
$\Omega=\sqrt{GM/R^3}$ \citep{lyndenbell74,pringle81}. We use a
one-dimensional layered disc model to follow the time evolution of the
total surface density, $\Sigma$, and midplane temperature, $T_{\rm c}$
as described by equations~1-8 in \cite{martin11a} but we do not assume
a constant surface density in the active layer. Instead, we use the
analytic approximations for the active layer surface density given in
equations~26 and~27 in MLLP1.  These are approximations to the
solution of the equation $Re_{\rm M}=Re_{\rm M,crit}$, assuming that
cosmic rays are the dominant source of ionisation. We note that this
description assumes Ohmic resistivity is the dominant non-ideal MHD
effect. However, recent works suggest that ambipolar diffusion and the
Hall effect may determine the active layer
\citep[e.g.][]{perezbecker11a,perezbecker11b,bai11}.  \cite{wardle11}
find the Ohmic resistivity term provides an average value for the
active layer surface density for a range of vertical magnetic
fields. Because Ohmic resistivity also has the advantage of being
independent of the magnetic field \citep{fleming00} we consider only
this effect here. We note that other sources of turbulence, such as the
baroclinic instability \citep[e.g.][]{klahr03,lyra11}, have been
suggested to generate viscosity within the dead zone. However, in
this work we we assume there is no turbulence within the dead zone.

In MLLP1 we found that the dead zone extent is not significantly
affected by variation in the metallicity and hence we use the high
metallicity limit for the electron fraction (see equation~19 in MLLP1)
and consider changes in the critical magnetic Reynolds number. The
active layer has a surface density $\Sigma_{\rm m}$, temperature
$T_{\rm m}$ and viscosity parametrised with the $\alpha$ prescription
by
\begin{equation}
\nu=\alpha c_{\rm m}H
\end{equation}
\citep{shakura73}, where the sound speed in the active layer is
\begin{equation}
c_{\rm m}=\sqrt{\frac{k T_{\rm m}}{\mu m_{\rm h}}},
\end{equation}
where $k$ is the Boltzmann constant and $\mu m_{\rm h}$ is the mean
molecular mass. We approximate the disc scale height by $H=c_{\rm
  m}/\Omega$ and so
\begin{equation}
\nu_{\rm m}=\alpha \frac{c_{\rm m}^2}{\Omega}.
\end{equation}
The \cite{shakura73} $\alpha$ parameter still has some
uncertainty. \footnote {We note that it may be more realistic to take
  $\nu=\alpha c_{\rm m} h$ where $h$ is the local scale height in the
  layer. The scale height in the active layer should be $h=\min (H,
  \rho/\frac{d\rho}{dz})$ at the base of the active layer and so
  typically $h\ll H$. As mass builds up in the dead zone, the
  magnetic surface layer will lie further above the disc midplane,
  resulting in higher local gravity and a smaller scale height. This
  means that the $\alpha$ parameter we take is effectively increased
  to $\alpha_{\rm eff}=(H/h)\alpha$. For example if we have
  $\alpha=0.01$ and $H/h=10$ then the effective parameter in the
  active layer is larger, $\alpha_{\rm eff}=0.1$. This should be
  investigated further in future work.}  Numerical MHD simulations find
$\alpha=0.01$ \citep{brandenburg95,stone96} but observations suggest
$\alpha \sim 0.1-0.4$ \citep{king07}.  If the disc is fully turbulent
the single layer has $\Sigma_{\rm m}=\Sigma$ and $T_{\rm m}=T_{\rm
  c}$. However, where a dead zone exists the disc has two layers and
the complementary region is defined with surface density $\Sigma_{\rm
  g}=\Sigma-\Sigma_{\rm m}$ and temperature $T_{\rm c}$. The dead
layer may become self-gravitating if $Q<Q_{\rm crit}$, where the
Toomre parameter is
\begin{equation}
Q=\frac{c_{\rm g}\Omega}{\pi G \Sigma},
\end{equation}
the sound speed at the midplane is
\begin{equation}
c_{\rm g}=\sqrt{\frac{k T_{\rm c}}{\mu m_{\rm h}}}
\end{equation} 
and we take $Q_{\rm crit}=2$.  Self-gravity drives a second effective
viscosity in the disc as explained in \cite{martin11a}.

In MLLP1, we found that the active layer surface density may be
significantly smaller than that assumed by the constant prescription,
especially at the inner edge of the dead zone. The active layer may
become optically thin and this affects the relationship between the
layer temperatures. The optical depth of the turbulent layer is
\begin{equation}
\tau_{\rm m}=\frac{3}{8} \kappa(T_{\rm m},\rho_{\rm m})\frac{\Sigma_{\rm m}}{2},
\end{equation}
where $\kappa$ is the opacity and the density is approximated as
$\rho_{\rm m}=\Sigma_{\rm m}/2H_{\rm m}$ with scale height $H_{\rm
  m}=c_{\rm m}/\Omega$. The optical depth within the complementary
region, if it exists, is
\begin{equation}
\tau_{\rm g}=\frac{3}{8} \kappa(T_{\rm c},\rho_{\rm c})\frac{\Sigma_{\rm g}}{2},
\end{equation}
where the density is $\rho_{\rm c}=\Sigma_{\rm g}/2H_{\rm g}$ and the
scale height is $H_{\rm g}=c_{\rm g}/\Omega$.  The total optical depth
of the disc is
\begin{equation}
\tau=\tau_{\rm m}+\tau_{\rm g},
\end{equation}
the sum of the optical depths of the two layers.

In this work we consider two different opacity prescriptions. The
first is a simple power law in temperature
\begin{equation}
\kappa(T)=0.02\,T^{0.8}\,\rm cm^2\,g^{-1}
\label{kappa}
\end{equation}
\citep{armitage01,martin11a}. This is useful for finding analytic
solutions because it does not depend on the density of the disc. It is
the opacity in the tables of \cite{bell94} around the temperature
where the MRI is triggered of around $800\,\rm K$
\citep[see][]{umebayashi83,gammie96}. The second is the full opacity
shown in table~1 of \cite{zhu09} that depends on both the temperature
and the density of the disc.

If the active layer is optically thick, the opacity $\tau_{\rm m} >1$, the
temperatures in the layers are found with
\begin{equation}
T_{\rm m}^4 =\tau_{\rm m} T_{\rm e}^4,
\label{one}
\end{equation}
where $T_{\rm e}$ is the surface temperature, and
\begin{equation}
\sigma \tau T_{\rm e}^4=  \sigma T_{\rm c}^4+\frac{9}{8}\nu_{\rm m}\Sigma_{\rm m}\Omega^2 \tau_{\rm g}.
\end{equation}
However, if the active layer is optically thin, $\tau_{\rm m}<1$, we
take $T_{\rm m}=T_{\rm e}$ and
\begin{equation}
\tau_{\rm g} T_{\rm e}^4 =  T_{\rm c}^4.
\end{equation}
The active layer may be optically thin in the inner parts of the disc
where the layer is thinnest.

With a large active layer surface density the inner parts of the disc
remain hot enough for thermal ionisation. However, when the dead zone
is determined by a critical magnetic Reynolds number the active layer
can become very thin and the inner parts of the disc provide little
accretional heating. In this case, heating from the star may dominate
the accretional heating. We consider this further in Section~\ref{irr}
but find it makes little difference except for radii $R\lesssim
0.1\,\rm AU$.

\section{Analytical Steady State Discs}
\label{steady}
 
For sufficiently small critical magnetic Reynolds numbers a steady
state disc solution exists that is fully turbulent, thus $\Sigma_{\rm
  m}=\Sigma$ everywhere.  There is no complementary region,
$\Sigma_{\rm g}=0$, and the temperatures are related with $T_{\rm
  m}=T_{\rm c}$ and
\begin{equation}
T_{\rm c}^4=\tau T_{\rm e}^4.
\label{temp}
\end{equation}
For a fully turbulent disc, the surface density is found with
\begin{equation}
\nu_{\rm m} \Sigma =\frac{\dot M}{3\pi} \left[1-\left(\frac{R_{\rm
      in}}{R}\right)^\frac{1}{2}\right]
\label{sig}
\end{equation}
\citep{pringle81}, where $\dot M=\dot M_{\rm infall}$ is the steady
accretion rate through the disc.  There is a zero torque inner
boundary condition applied at $R=R_{\rm in}=5\,\rm R_\odot$ where the
mass falls freely on to the star. Similarly, the surface temperature
is found with
\begin{equation}
\sigma T_{\rm e}^4=\frac{3 \dot M \Omega^2}{8\pi
}\left[1-\left(\frac{R_{\rm in}}{R}\right)^\frac{1}{2}\right]
\label{te}
\end{equation}
\citep{pringle81}.  In this section, we use the simplified power law
opacity given in equation~(\ref{kappa}) so that we can easily solve
equations~(\ref{temp}),~(\ref{sig}) and~(\ref{te}) analytically for
the surface density and temperature in the disc.  With this fully
turbulent disc solution, we check if the solution has a dead zone (we
solve $Re_{\rm M}=Re_{\rm M,crit}$). If it does, then the steady state
does not exist and we expect time dependent accretion (see
Section~\ref{time}).

In Fig.~\ref{layer} we show the steady state surface density of the
disc around a star of mass $M=1\,\rm M_\odot$ for varying infall
accretion rates with critical magnetic Reynolds number $Re_{\rm
  M,crit}=10^3$ and $10^4$ with $\alpha=0.1$ on the left and
$\alpha=0.01$ on the right.  For larger $\alpha$, the steady state
surface density is smaller. Unless the accretion rate is very small,
there is no dead zone for $Re_{\rm M,crit}\lesssim 10^4$.

\begin{figure}
\includegraphics[width=8.9cm]{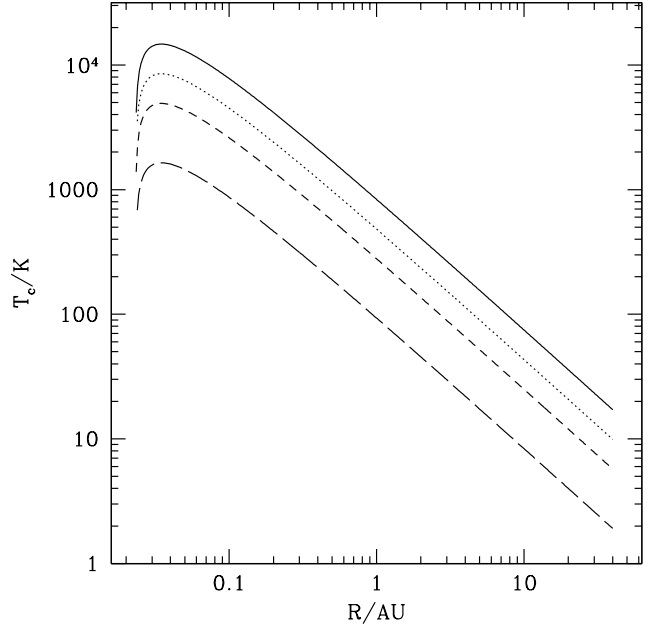}
\caption{ The temperature of the steady state discs shown in
  Fig.~\ref{layer} with $\alpha=0.1$ for $\dot M_{\rm infall}=10^{-6}\,\rm
  M_\odot\,yr^{-1}$ (solid line), $\dot M_{\rm infall}=10^{-7}\,\rm
  M_\odot\,yr^{-1}$ (short-dashed line) and $\dot M_{\rm infall}=10^{-8}\,\rm
  M_\odot\,yr^{-1}$ (long dashed line). The dotted line shows
  $\alpha=0.01$ and $\dot M_{\rm infall}=10^{-7}\,\rm M_\odot\,yr^{-1}$.}
\label{fig:temp}
\end{figure}

In Fig.~\ref{fig:temp} we show the corresponding disc temperatures,
for the fully turbulent disc, that decrease with the infall accretion
rate. Because the active layer surface density decreases with
temperature (MLLP1) the active layer surface density also decreases
with the infall accretion rate. In a time dependent disc, this steady
solution can only exist if the disc is fully turbulent. 

In MLLP1 we found that the active layer surface density is only
sensitive to the temperature and not the total surface density. The
temperature in the steady state disc found here is higher than often
considered in static discs \citep[e.g.][]{fromang02,matsumura03}.  The
new result is that for time evolving discs, dead zones exist only for
large critical magnetic Reynolds number. In order to investigate discs
with large critical magnetic Reynolds numbers, and hence dead zones,
we must solve the time-dependent accretion disc equations.

\section{Time-Dependent Simulations}
\label{time}

We solve numerically the time-dependent layered accretion disc
equations with a one-dimensional code similar to \cite{martin11a}. The
disc around a star of mass $M=1\,\rm M_\odot$ extends from $R_{\rm
  in}=5\,\rm R_\odot$ up to $R_{\rm out}=40\,\rm AU$ \citep[see
  e.g.][]{armitage01,martin11a,martin11b}. The grid has 200 points
distributed uniformly in $\log R$ so that at each radius the grid
point separation is $\Delta R/R =0.037$.  Material is continuously
added to the disc at a radius of $R_{\rm add}=35\,\rm AU$ at a rate
$\dot M_{\rm infall}$. The inner edge of the disc has a zero torque
inner boundary condition and the outer edge has a zero radial velocity
boundary condition that prevents outward flow. The initial conditions
do not have any effect on the outcome because we evolve the disc until
it reaches a steady state or a limit cycle in the accretion rate.  We
consider both the simple power law opacity given in
equation~(\ref{kappa}) and also a full opacity given in table~1 of
\cite{zhu09} in our numerical models.

\begin{figure*}
\includegraphics[width=8.4cm]{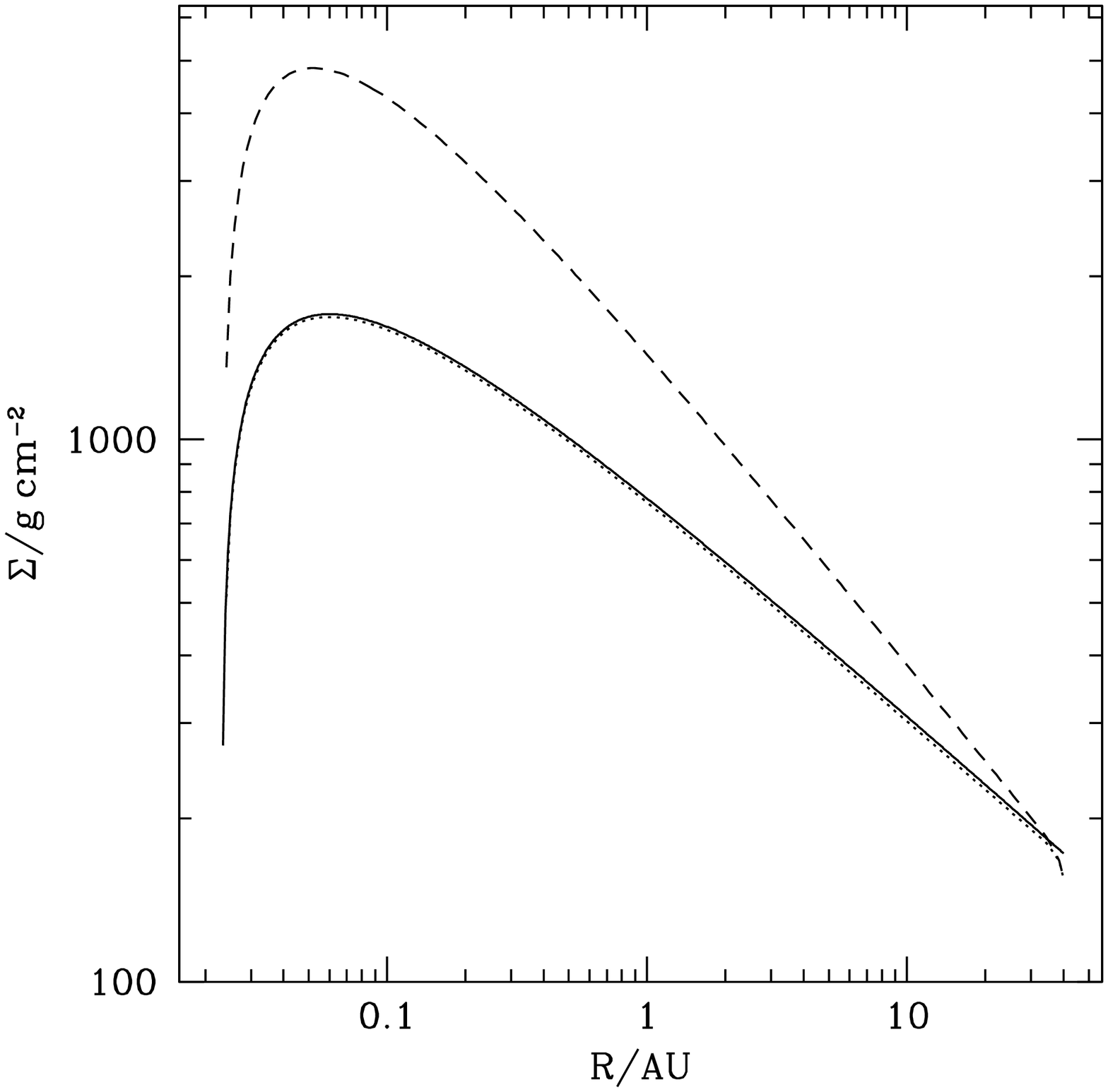}
\includegraphics[width=8.4cm]{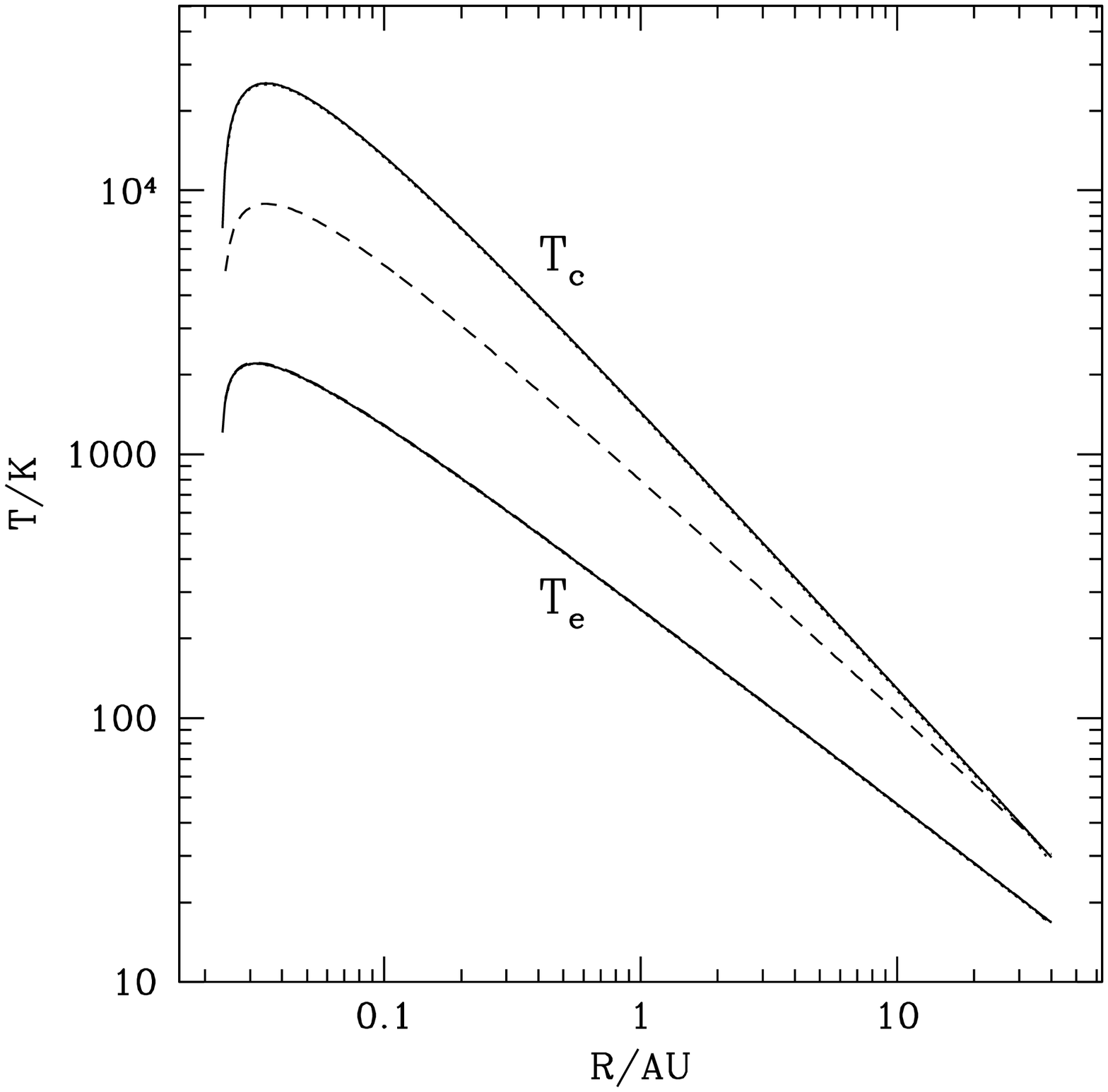}
\caption{The fully turbulent steady state disc surface density (left)
  and temperatures (right) with an infall accretion rate $\dot M_{\rm
    infall}=10^{-6}\,\rm M_\odot\,yr^{-1}$ and $\alpha=0.01$. The
  solid lines show the analytical model given in Section~\ref{steady}
  with the simple opacity law. The dotted lines show the corresponding
  numerical model with the simple opacity law (this is almost
  identical to the analytical solution).  The dashed lines show the
  numerical model with the full opacity table. For the temperature
  plot we show the midplane temperature, $T_{\rm c}$, (upper three
  lines) and surface temperature, $T_{\rm e}$ (lower three lines). The
  surface temperatures are almost identical for all three models. }
\label{fig:surf2}
\end{figure*}

\subsection{Small Critical Magnetic Reynolds Number}
\label{low}

Only with large critical magnetic Reynolds numbers does a dead zone
exist in the steady state disc. However, our analytical models in the
previous section used only the simple power law opacity. In this
Section, we numerically evolve the disc until it reaches a steady
state (with no dead zone) in order to examine properties of a disc
with the full opacity prescription.

In Fig.~\ref{fig:surf2} we show the surface density of the fully
turbulent steady state disc with an infall accretion rate of $\dot
M_{\rm infall}=10^{-6}\,\rm M_\odot\,yr^{-1}$ and $\alpha=0.01$.  The
simple power law opacity approximates well the analytical solutions
shown.  In the outer parts of the disc, the numerical solutions have a
zero radial velocity condition imposed that alters the solution
slightly from the analytical one. We also show a simulation for a disc
with the full opacity table. The inner parts of the disc are
significantly more massive than with the simple opacity law. However,
these parts are hot enough to be thermally ionised.

In Fig.~\ref{fig:surf2} we also show the midplane and surface
temperatures of the steady state disc. The numerical simulation with
the simple opacity law approximates well the analytical steady state
solution. The simulation with the full opacity table shows that the
inner parts of the disc may be slightly cooler compared with the
simple opacity law. However, further out in the disc where dead zone
formation occurs the power law is a reasonable approximation to the
full opacity. Hence, even though we have taken a simple opacity law in
the previous section, it does not change our conclusion significantly
that dead zones will only form for large critical magnetic Reynolds
numbers $Re_{\rm M,crit} \gtrsim 10^4$ unless the accretion rate is
very small (we have also verified this numerically).

\subsection{Large Critical Magnetic Reynolds Number}

For sufficiently large critical magnetic Reynolds number, the steady
state disc solution contains a dead zone. Hence, it cannot be in
steady state and we expect the disc to be unstable to the
gravo-magneto instability. This results in outbursts in the accretion
rate on to the central object \citep{gammie96,armitage01,zhu09}. We
use the analytic approximation for the active layer surface density
given in equations~26 and~27 in MLLP1 that is valid for $Re_{\rm
  M,crit}\gtrsim 100$.  We choose $Re_{\rm M,crit}=10^4$, $\dot M_{\rm
  infall}=10^{-6}\,\rm M_\odot\,yr^{-1}$ and $\alpha=0.01$ so that we
can compare with previous work with a constant surface density in the
active layer. We use the simple power law opacity because we have seen
in Section~\ref{low} that it makes little difference except to the
hottest parts of the disc which are already thermally ionised and
fully turbulent.

\begin{figure}
\includegraphics[width=8.9cm]{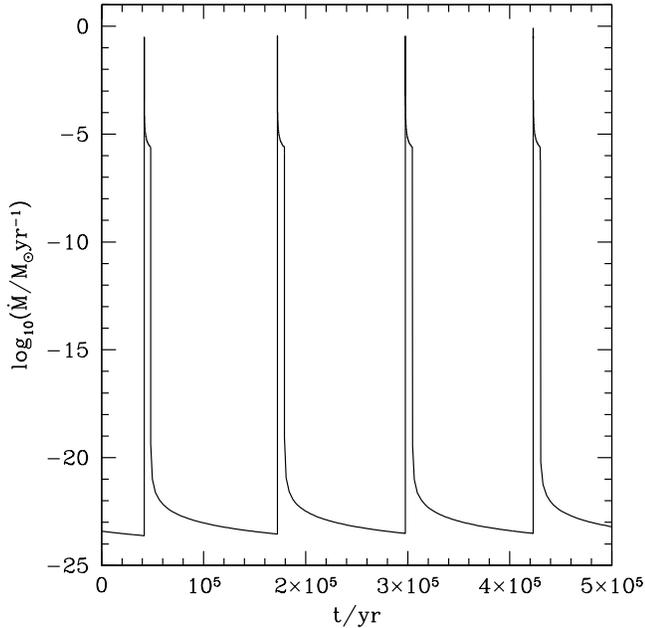}
\caption{The accretion rate on to the solar mass central star for a
  disc with $\dot M_{\rm infall}=10^{-6}\,\rm M_\odot\,\rm yr^{-1}$,
  $\alpha=0.01$ and $Re_{\rm M,crit}=10^4$. A mass of $0.1\,\rm
  M_\odot$ is accreted on to the central star during each outburst. }
\label{acc}
\end{figure}

\begin{figure}
\includegraphics[width=8.9cm]{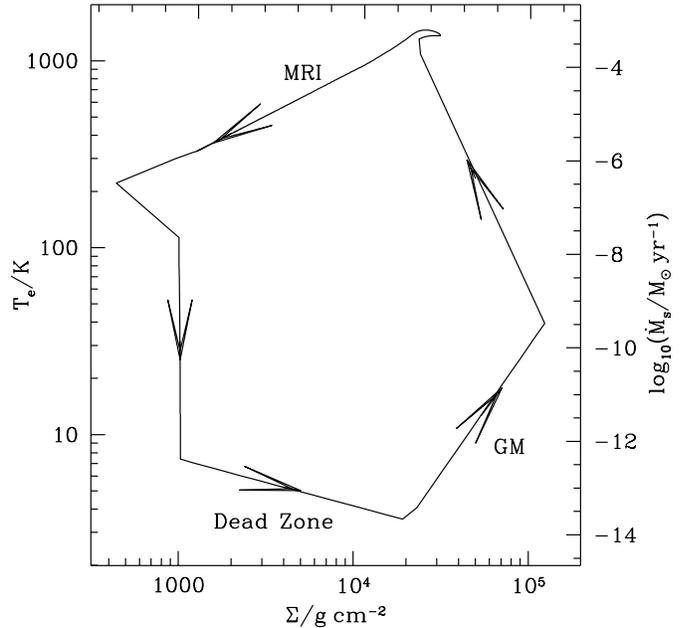}
\caption{The limit cycle in the $\Sigma-\dot M_{\rm s}$ or
  $\Sigma-T_{\rm e}$ plane at $R=1\,\rm AU$ for the disc in
  Fig.~\ref{acc}. The disc moves in an anti-clockwise direction around
  the cycle. The steady accretion rate $\dot M_{\rm s}$ corresponds to
  the steady state surface temperature given in equation~(\ref{te}). }
\label{limit}
\end{figure}

In Fig.~\ref{acc} we show the accretion rate on to the central
star. When the dead zone is present, the accretion rate on to the star
is extremely low. The accretion outbursts occur on a timescale of
$1.1\times 10^5\,\rm yr$ and a mass of around $0.1\,\rm M_\odot$ is
accreted during each.  After a few outbursts the disc reaches a limit
cycle that repeats itself.  Following \cite{martin11a} and
\cite{lubow12}, in Fig.~\ref{limit} we show the limit cycle in the
$\Sigma-\dot M_{\rm s}$ plane at a radius of $R=1\,\rm AU$. The steady
accretion rate $\dot M_{\rm s}$ corresponds to the steady surface
temperature in equation~(\ref{te}). The disc moves in an
anti-clockwise direction around the cycle. In the dead zone branch,
the active layer deposits material on to the dead zone and as a
result, the surface density increases along the branch. The accretion
rate decreases with increasing surface density because the active
layer decreases. The disc becomes self-gravitating along the
gravo-magneto (GM) branch. The temperature, and hence accretion rate,
begins to rise as the surface density continues to increase.  The disc
eventually becomes hot enough for thermal ionisation to take place at
some radius and the MRI is triggered at the end of the GM branch. The
disc moves quickly to the upper MRI branch. Now the accretion rate on
to the star is much larger than the accretion on to the disc and so
the accretion rate declines as the surface density decreases. The dead
zone re-forms, the disc quickly transitions down to the dead zone
branch and the cycle repeats.

\begin{figure*}
\includegraphics[width=8.4cm]{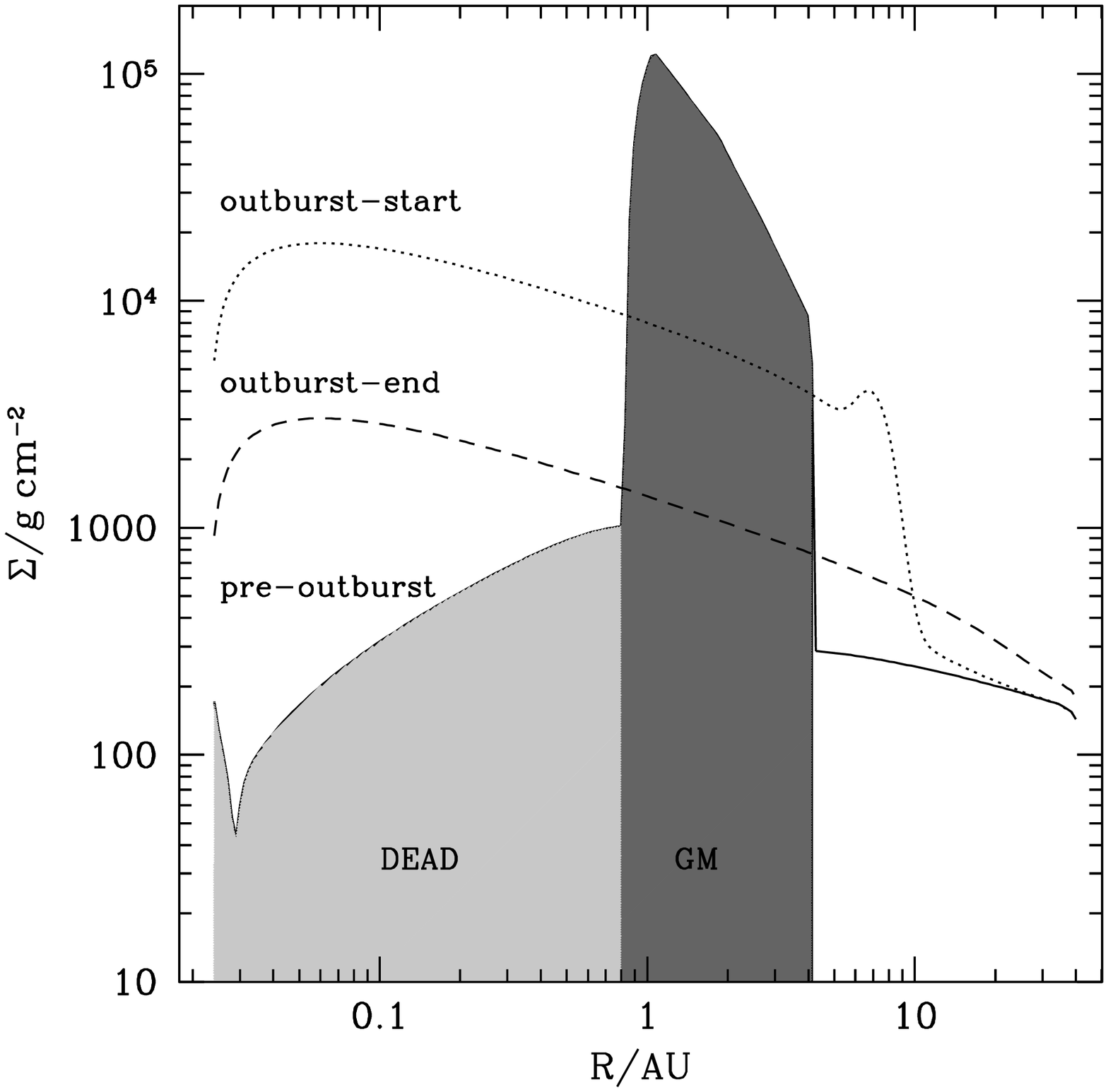}
\includegraphics[width=8.4cm]{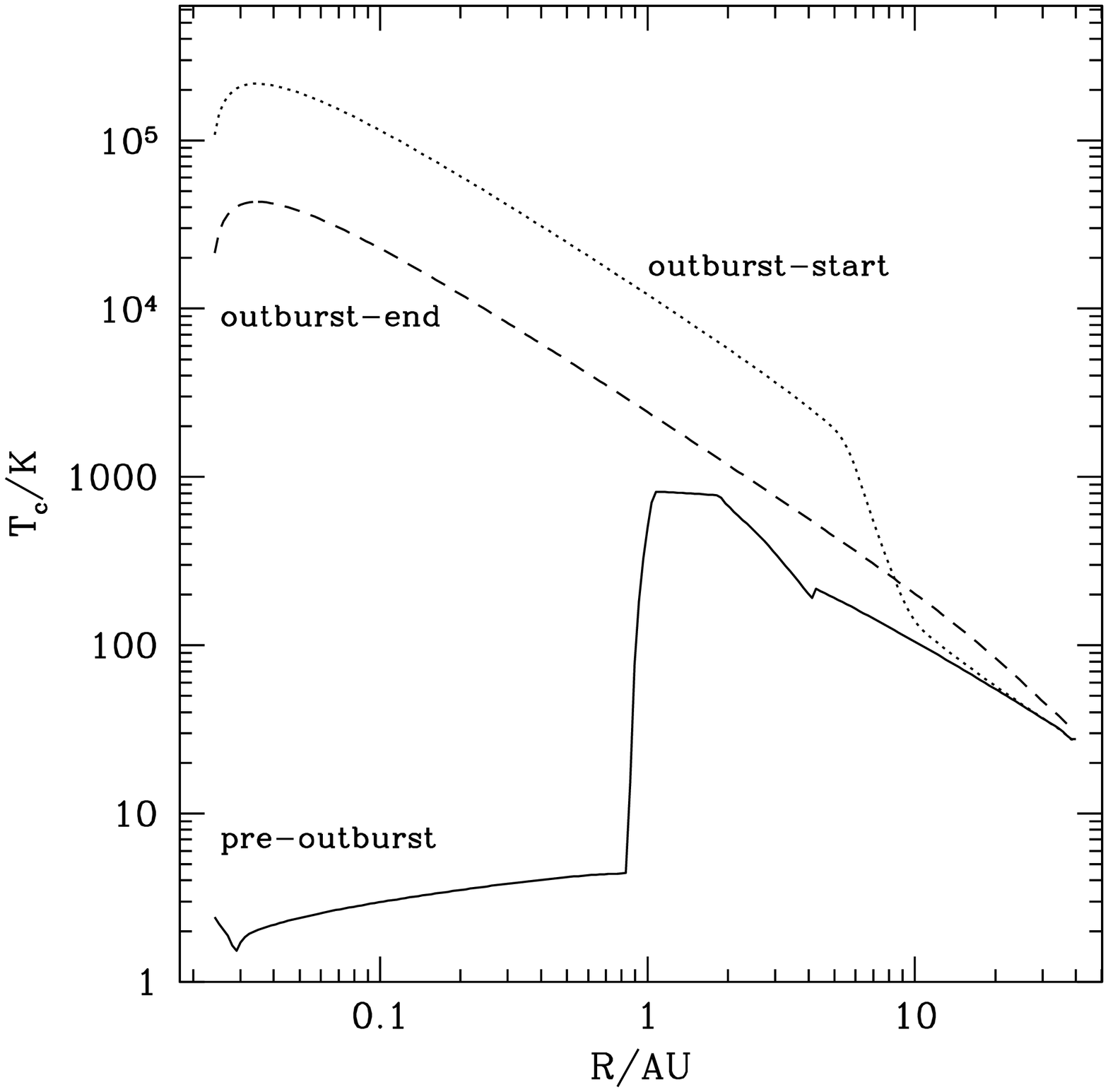}
\caption{Numerical simulations with $M=1\,\rm M_\odot$, $\dot M_{\rm
    infall}=10^{-6}\,\rm M_\odot\,yr^{-1}$, $Re_{\rm M,crit}=10^4$ and
  $\alpha=0.01$. The surface density (left) and midplane temperature
  (right) in the disc just before an outburst (solid lines and shaded
  region), at the start of an an outburst (dotted lines) and towards
  the end of an outburst (dashed lines).  Just before the outburst,
  the dead zone is shown in the shaded region.  The darker shaded part
  (labeled GM) shows where the dead zone is self-gravitating.  The
  upper dead zone layer boundary is too close to the total surface
  density to be seen to be different on this scale (the active layer
  is very thin). During the outburst the whole disc is turbulent.  }
\label{sd}
\end{figure*}

In Fig.~\ref{sd} we show the surface density and midplane temperature
of the disc just before an outburst and during an outburst. There is
very little mass in the inner parts of the accretion disc. This is
different to the disc with a constant (and large) surface density in
the active layer.  The active layer here is very thin and little mass
flows around the dead zone. Material in the dead zone builds up and
peaks in the disc where it becomes self-gravitating. The temperatures
also peak at this radius. The MRI is first triggered at the peak and
the outburst begins. We have found that the gravo-magneto instability
mechanism is the same no matter how the extent of the dead zone is
determined.

\subsection{Irradiation from the star}
\label{irr}

When the active layer is very thin, in the innermost parts of the disc
the accretional heating may be dominated by the heating from the
star. The flux of radiation is
\begin{equation}
F_{\rm irr}= \sigma T_\star^4 \,\frac{\alpha_{\rm
    irr}}{2}\left(\frac{R_\star}{R}\right)^\frac{1}{2}
\end{equation}
where
\begin{equation}
\alpha_{\rm irr}=0.005\left(\frac{R}{\,\rm
  AU}\right)^{-1}+0.05\left(\frac{R}{\rm AU}\right)^\frac{2}{7}
\end{equation}
\citep{chiang97} and we take the radius of the star to
be $R_\star=5\,\rm R_\odot$ and the temperature $T_\star=4000\,\rm K$.
If the midplane temperature of the disc drops below the irradiation
temperature
\begin{equation}
T_{\rm irr}=\left(\frac{F_{\rm irr}}{\sigma}\right)^\frac{1}{4},
\end{equation}
then we assume that the disc is isothermal and set $T_{\rm c}=T_{\rm
  m}=T_{\rm e}$. 

When we include this term the inner parts of the disc are warmer and
as a result, up to a radius of around $0.1\,\rm AU$ the disc is
thermally ionised. However, because the innermost part of the dead
zone still has a very small active layer, the accretion rate on to the
star remains small. For example, for the model in the previous
section, without the irradiation term we find that the accretion rate
between outbursts is around $10^{-25}\,\rm M_\odot\,yr^{-1}$ whereas
with the effects of irradiation this increases to around
$10^{-16}\,\rm M_\odot\,yr^{-1}$. This is still significantly too
small to account for the T Tauri accretion rates that are thought to
be around $10^{-8}\,\rm M_\odot\,yr^{-1}$ at an age of $1\,\rm Myr$
\citep{valenti93, hartmann98}. For the rest of this work we ignore the
heating from the star because is only affects a very small part of the
disc.

\section{FU Orionis Outbursts}
\label{dec}

In a collapsing molecular cloud the accretion rate on to the disc will
not be constant in time. We follow \cite{armitage01} who use a simple
model for the the protostellar accretion history. At early times the
infall rate is expected to be $\dot M_{\rm infall}=c_{\rm s}^3/G$
where $c_{\rm s}$ is the sound speed in the cloud \citep{shu77}. For a
temperature of $10\,\rm K$, this is around $\dot M_{\rm infall}\approx
10^{-5}\,\rm M_\odot\,yr^{-1}$. We assume that the accretion rate
declines exponentially in time
\begin{equation}
\dot M_{\rm infall}=\dot M_{\rm i}\exp\left(-\frac{t}{t_{\rm ff}}\right),
\label{declineq}
\end{equation}
where we take the initial accretion rate to be $\dot M_{\rm i}=2\times
10^{-5}\,\rm M_\odot\,yr^{-1}$. At this time, we begin with an
initially fully turbulent steady state accretion disc. The free fall
timescale is
\begin{equation}
t_{\rm ff}=\left(\frac{3\pi}{32 G \rho_{\rm cloud}}\right)^\frac{1}{2},
\end{equation}
where $\rho_{\rm cloud}$ is the cloud density. We take $t_{\rm
  ff}=10^5\,\rm yr$ \citep{armitage01}.

\begin{figure}
\includegraphics[width=8.4cm]{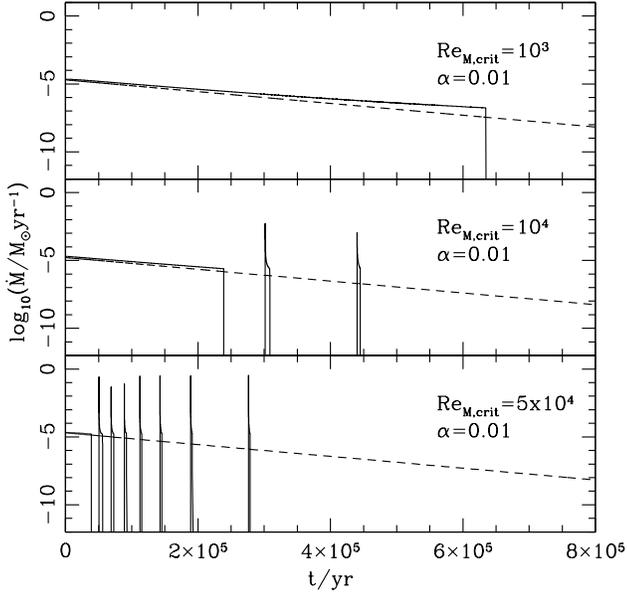}
\caption{The evolution of the accretion rate on to a solar mass
  star. The dashed lines show the exponentially decreasing infall
  accretion rate given in equation~(\ref{declineq}). The solid lines
  show the accretion rate on to the star with $\alpha=0.01$, $Re_{\rm
    M,crit}=10^3$ (top), $10^4$ (middle) and $5\times 10^4$
  (bottom). Initially the infall accretion rate is high enough that
  there is no dead zone. The accretion rate on to the star proceeds at
  a rate similar to that on to the disc. As the infall accretion rate
  drops, the temperature decreases and a dead zone forms. Then the
  accretion proceeds in large outbursts with a very small rate in
  between. At late times there is not sufficient material left in the
  disc for outbursts to occur. When the critical magnetic Reynolds
  number is $10^3$, there is not sufficient mass left in the disc for
  outbursts by the time the dead zone has formed.}
\label{decline}
\end{figure}

With a star formation rate predicted by \cite{miller79},
theoretically, \cite{hartmann85} predict that each FU Orionis system
should show around ten outbursts during its lifetime. In
Fig.~\ref{decline} we show the accretion rate on to the central star
as a function of time for the infall accretion rate on to the disc
given in equation~(\ref{declineq}) with $\alpha=0.01$ and three
different critical magnetic Reynolds numbers. In all cases, initially
the disc is fully turbulent and the accretion on to the star proceeds
at a rate close to $\dot M_{\rm infall}$. However, once a dead zone
forms, the accretion rate drops. For a solar mass star with a disc
with $\alpha=0.01$, the critical magnetic Reynolds number must be of
the order of a few times $10^4$ for a significant number of outbursts
to occur. For the lowest critical magnetic Reynolds number shown of
$10^3$, a dead zone forms once the infall accretion rate has dropped
to about $3\times 10^{-8}\,\rm M_\odot\,yr^{-1}$. However, at this
time there is not enough material in the disc for outbursts to occur. 

\begin{figure}
\includegraphics[width=8.4cm]{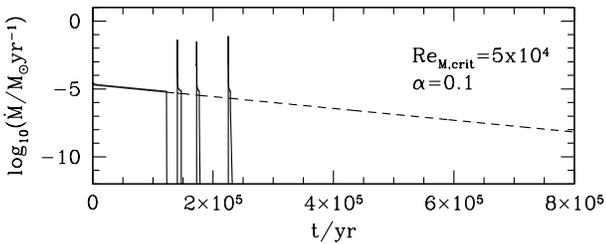}
\caption{As Fig.~\ref{decline} but with $\alpha=0.1$ and $Re_{\rm
    M,crit}=5 \times 10^4$. }
\label{decline2}
\end{figure}

In Fig.~\ref{decline2} we show the accretion rate on to the star for
$\alpha=0.1$ and $Re_{\rm M,crit}=5\times 10^4$. For a higher $\alpha$
we need an even larger critical Reynolds number for outbursts to
occur. We note that the evolution is also dependent on the mass of the
star. The active layer surface density decreases as the mass of the
central star increases (MLLP1). Increasing the mass of the central
star has the same effect as increasing the critical magnetic Reynolds
number. So, for larger mass stars, dead zones will form for a smaller
critical magnetic Reynolds number.

If the observed FU Orionis outbursts are driven by the gravo-magneto
instability, then the critical magnetic Reynolds number must be very
large, of order a few $10^4$. The outbursts are similar to those
previously found with a model with constant surface density in the
active layer. However, we predict a very small accretion rate in
between outbursts.  We should try to observe FU Orionis systems in
quiescence in order to draw conslusions about the surface density in
the active layer.  It is possible that effects such as ambipolar
diffusion or the Hall effect could play a role in increasing the
active layer surface density in the inner regions and this should be
investigated in future work.

\section{Transition Discs}
\label{trans}

At later times of disc evolution, the infall accretion rate on to the
disc slows and the mass and temperature of the disc both decrease. A dead
zone determined by a critical magnetic Reynolds number becomes more
prominent as the disc cools (e.g. see Fig.~1). However, with the
constant active layer surface density previously assumed this is not
the case.  The constant active layer model predicts that when the
surface density drops below the critical value (of around $200\,\rm
g\,cm^{-2}$) the whole disc becomes turbulent.   The
increasingly large dead zone found with a critical magnetic Reynolds
number has implications for low mass discs in the later stages of
evolution.

After the infall on to the disc ends, there is not sufficient
accretion through the disc for it to build up enough material for
accretion outbursts (see for example top pannel in Fig.~\ref{decline}). We
have investigated this numerically and find the surface density and
temperature distributions are similar to the pre-outburst distribution
shown in Fig.~\ref{sd}. The inner parts of the disc, that are mostly
dead, are depleted by several orders of magnitude compared with the
outer parts that are self-gravitating. Similarly the temperature is
several orders of magnitude smaller in the inner regions. Only a small
amount of material can flow through the thin active layer on top of
the dead zone. The depleted inner parts of the disc could appear to
form a hole at the inside on the viscous timescale
\begin{align}
\tau_{\nu} & =  \frac{R^2}{\nu}  \cr
& = 1.41\times 10^4\, \left(\frac{\alpha}{0.01}\right)^{-1} \left(\frac{T}{100\,\rm K}\right)^{-1}
\left(\frac{M}{1\,\rm M_\odot}\right)^\frac{1}{2}\cr
&\,\,\,\,\,\,\, \times \left(\frac{R}{1\,\rm AU}\right)^\frac{1}{2} \,\rm yr.
\end{align}
This is very short compared with the lifetime of the disc.

A large dead zone could help to explain observed transition objects
without the need for a photoevaporation model. The surface density
distribution is qualitatively similar to that shown in Figure~1 in
\cite{clarke01} for their ultra violet switch photoevaporation
model. However, that model takes a few $10^6$ years to form the inner
hole whereas the dead zone model forms a hole much more quickly. We
have not included dust grains in our models. These suppress the
ionisation of the disc further leading to a larger dead zone
\cite[e.g.][]{bai09,perezbecker11a,bai11}.  The dead zone model should
be investigated further in future work with respect to observations of
transtion discs.

\section{Conclusions}

We have presented global time-dependent calculations of circumstellar
discs with a dead zone determined by a critical magnetic Reynolds
number.  For infall accretion rates around $10^{-7}\,\rm
M_\odot\,yr^{-1}$, if the critical magnetic Reynolds number is larger
than about $10^4$, the disc around a young stellar object forms a dead
zone. If the critical magnetic Reynolds number is smaller, then the
disc is fully turbulent.

If the infall accretion rate is sufficiently high when a dead zone
forms, the disc becomes unstable to the gravo-magneto instability.
The outburst behaviour is similar to that found previously by assuming
a constant active layer surface density. If the gravo-magneto
instability is responsible for FU Orionis outbursts, then the critical
magnetic Reynolds number must be greater than a few $10^4$.

A disc model with a dead zone determined by a critical magnetic
Reynolds number predicts a very low accretion rate between outbursts
that was not predicted with the constant active layer surface density
assumption.  This low quiescent rate should be observed both before
and after FU Orionis outbursts take place. Alternatively, the
accretion rate the model predicts may be increased if effects such as
ambipolar diffusion or the Hall effect are included.

The size of a dead zone in a disc increases as the disc cools. We
suggest that large inner dead zones may help to explain observations
of the depleted inner regions of transition discs. However, there are
some limitations to the model used here. Some of these have already
been discussed in \cite{martin11a} and MLLP1. There is much more work
to be done in future to establish criteria for the formation of dead
zones. This is not straightforward because it depends on assumptions
about the magnetic field and the still unresolved problem of the
viscosity $\alpha$ parameter. Once the criteria are established, the
evolution of a protostellar disc should be computed, from times of
high infall rates to late time disc dissipation.

\section*{Acknowledgements}

RGM thanks the Space Telescope Science Institute for a Giacconi
Fellowship. SHL acknowledges support from NASA grant NNX07AI72G. JEP
thanks the Collaborative Visitor Program at STScI for its support and
hospitality


\label{lastpage}
\end{document}